**PREPRINT (not peer-reviewed)**

**TITLE:** ESSENTIAL REQUIREMENTS FOR ESTABLISHING AND OPERATING DATA TRUSTS: PRACTICAL GUIDANCE BASED ON A WORKING MEETING OF FIFTEEN CANADIAN ORGANIZATIONS AND INITIATIVES

P. Alison Paprica, University of Toronto, Institute of Health Policy, Management and Evaluation; Vector Institute; Health Data Research Network Canada; ICES alison.paprica@utoronto.ca (corresponding author); Eric Sutherland, Canadian Institute for Health Information, esutherland@cihi.ca; Andrea Smith, Vector Institute, andrea.smith@vectorinstitute.ai; Michael Brudno, HPC4Health; University Health Network; Hospital for Sick Children; University of Toronto, Department of Computer Science, brudno@cs.toronto.edu; Rosario G. Cartagena, ICES, rosario.cartagena@ices.on.ca; Monique Crichlow, Compute Ontario, monique.crichlow@computeontario.ca; Brian K Courtney, Sunnybrook Research Institute, brian.courtney@sunnybrook.ca; Chris Loken, Compute Ontario, cloken@computeontario.ca; Kimberlyn M. McGrail, Population Data BC; Centre for Health Services and Policy Research, University of British Columbia, Faculty of Medicine, School of Population and Public Health; Health Data Research Network Canada, kim.mcgrail@ubc.ca; Alex Ryan, MaRS Discovery District, alexryan@marsdd.com; Michael J Schull, ICES; Health Data Research Network Canada, michael.schull@ices.on.ca; Adrian Thorogood, McGill University, Centre of Genomics and Policy; Global Alliance for Genomics and Health, adrian.thorogood@mcgill.ca; Carl Virtanen, University Health Network, HPC4Health, carl.virtanen@uhnresearch.ca; Kathleen Yang, Canadian Institute for Health Information, kayang@cihi.ca

## ABSTRACT

**Introduction:** Increasingly, the label "data trust" is being applied to repeatable mechanisms or approaches to sharing data in a timely, fair, safe and equitable way. However, there is a gap in terms of practical guidance about how to establish and operate a data trust.

**Aim and Approach:** In December 2019, the Canadian Institute for Health Information and the Vector Institute for Artificial Intelligence convened a working meeting of 19 people representing 15 Canadian organizations/initiatives involved in data sharing, most of which focus on public sector health data. The objective was to identify essential requirements for the establishment and operation of data trusts. Preliminary findings were presented during the meeting then refined as participants and co-authors identified relevant literature and contributed to this manuscript.

**Results:** Twelve (12) minimum specification requirements ("min specs") for data trusts were identified. The foundational min spec is that data trusts must meet all legal requirements, including legal authority to collect, hold or share data. In addition, there was agreement that data trusts must have (i) an accountable governing body which ensures the data trust advances its stated purpose and is transparent, (ii) comprehensive data management including responsible parties and clear processes for the collection, storage, access, disclosure and use of data, (iii) training and accountability requirements for all data users and (iv) ongoing public and stakeholder engagement.

**Conclusion / Implications:** Based on a review of the literature and advice from participants from 15 Canadian organizations/initiatives, practical guidance in the form of twelve min specs for data trusts were agreed on. Public engagement and continued exchange of insights and experience is recommended on this evolving topic.

# BACKGROUND

Organizations around the world are actively working on ways to increase uses of person-level data for research, evaluation, planning and innovation while ensuring that data are secure, and privacy is protected [1-6]. These activities can be understood to be part of a broader effort to ensure appropriate data governance and management at a time when there is unprecedented potential to transform data into beneficial knowledge, but also high sensitivity and public concern about how data are shared, protected and used [7-12].

Over the past decade of digitization of records and services, organizations have learned how to govern and manage internal data, including personal information, to keep data secure and protect privacy. Organizations have also learned how to de-identify, aggregate and share data with low potential for misuse as open data. Between the extremes of internal data (including proprietary data) and open data, there remains a practical gap: how to responsibly share data and provide access where there is both a clear public benefit and a potential for misuse?

The term "data trust" received heightened attention when it was identified as a key mechanism to grow artificial intelligence (AI) in the UK in the 2017 Hall-Presenti report.[13] The report emphasizes the need for terms and mechanisms to facilitate the sharing of data between organizations that hold data (data providers) and organizations seeking to use data (data users). The Hall-Presenti report is direct in stating that the data trusts it envisions are "not legal entities or institutions, but rather a set of relationships underpinned by a repeatable framework, compliant with parties' obligations, to share data in a fair, safe and equitable way" [13].

There are many other working definitions of data trusts, some of which directly contradict the Hall-Presenti report's position that data trusts are not legal entities. For example, in 2020 the Open Data Institute put forward a working definition which draws upon the concept of a legal trust with trustees and beneficiaries: "a data trust provides independent, fiduciary stewardship of data" [14]. In addition, there are myriad additional labels applied to endeavors to responsibly share and provide access to data including: digital trusts, data co-operatives, data commons, data clubs, data institutions, data banks, data stewardships, data collaboratives and data safe havens [15-27].

One of the negative effects of the multiple labels and conflicting definitions is that it can obscure commonalities behind approaches to data sharing and data access. For example, the authors of this report have, at times, used several of the labels above to describe our work, while having common aspirations related to data that are FAIR (findable, accessible, interoperable, reusable) [28], and well governed and managed as per the Five Safes [29] and other frameworks [27,30,31]. Given recent large scale Canadian public investment in data infrastructure with an initial focus on data from publicly-funded health services, our group identified a need for practical guidance that goes beyond labels and focuses on how to establish and operate data infrastructure which supports data sharing and enables access to data while continuing to ensure data protection. Since our focus was not on exclusive definitions, we modified the Hall-Presenti report language and used the working definition "a data trust is a repeatable mechanism or approach to sharing data in a timely, fair, safe and equitable way" which neither requires nor precludes the data trust taking the form of a legal entity or independent institution. Our aim was to combine first-hand experience establishing data infrastructure with a synthesis of concepts in the literature in order to develop a common understanding of the essential requirements for data trusts, irrespective of the form that a data trust may take.

# METHOD

The Data Trust Working Meeting was the first "Capability Exchange" organized under CIHI's *Health Data and Information Capability Framework* which includes "facilitating exchange of knowledge" and "exploring harmonization" as two of its objectives. Participants were invited based on their organizations' active work on accessible data in Canada. In total 19 people representing 15 organizations and data infrastructure initiatives participated. Most participants work at publicly funded organizations focused on health data and/or data associated with publicly funded services. Several participants were involved in more than one initiative or organization, including some commercial organizations. However, to mitigate the risk that a single company could have a disproportionate influence on discussions and the outputs of the meeting, invites were not extended to representatives from commercial data sharing initiatives, though it was acknowledged that companies could have multiple roles to play in data trusts including as data providers and as leaders in technology-based approaches to monitoring, security, data governance and/or provenance.

Each participating organization/initiative was asked to provide a written summary of its activities which was circulated in advance of the meeting. We held a six-hour in-person meeting in Toronto on December 3, 2019 which comprised brief (~5 minute) presentations about each organization/initiative, followed by a series of facilitated discussions. The meeting utilized a minimum specifications requirements ("min specs") approach to identify the essential elements and key characteristics of data trusts [32]. This entailed inviting individuals to brainstorm a list of elements and characteristics that might be essential for data trusts and then, as a group, determining which should be crossed off the list based on the fact that it could be possible to have a complete and well-functioning data trust without them. Live internet polling was used to capture individual suggestions and key points from the group discussions. Preliminary findings were presented during the meeting then refined as co-authors contributed to this manuscript and identified relevant literature.

# RESULTS

Box 1: Min Specs for Data Trust Establishment and Operations

---

1. Legal: The data trust must fulfill all legal requirements, including the authority to collect, share and hold data
2. Governance
   a) The data trust must have a stated purpose
   b) The data trust must be transparent in its activities
   c) The data trust must have an accountable governing body
   d) Governance must be adaptive
3. Management
   a) There must be well-defined policies and processes for the collection, storage, use and disclosure of data
   b) Policies and processes must include data protection safeguards which are reviewed and updated regularly
   c) There must be an ongoing process to identify, assess and manage risks
4. Data User Requirements
   a) All data users must complete training before they access data
   b) All data users must agree to a data user agreement which acknowledges that data use will be monitored and includes consequences for non-compliance
5. Public and Stakeholder Engagement
   a) There must be early and ongoing engagement with stakeholders including members of the public
   b) Where there is a reasonable expectation that specific subpopulations or groups would have a particular interest in, or would be affected by, an activity of the data trust, there must be direct engagement tailored for that subpopulation/group

---

## Requirement 1: Legal – One (1) Min Spec

The foundational minimum specifications requirement (min spec "1") is that a data trust must fulfill all legal requirements. Organizations contemplating establishing and/or being part of a data trust need to be fully aware of, and be able to comply with, relevant legislation and regulations, for example around collecting, using and disclosing personal information. In Canada, private-sector organizations that collect, use or disclose personal information in the course of a commercial activity must comply with PIPEDA [33]. Similar data sharing activities for public sector data must comply with provincial legislation related to privacy, such as PHIPA in Ontario [34] and FIPPA in British Columbia [35]. For that reason, fulfilling this min spec may be more complex for cross-border data trusts, as they will need to identify multiple legal requirements and ensure that governance addresses all of them.

In addition to legislation and regulations, there are typically binding terms and conditions in data sharing agreements established between legal entities when data are shared (e.g., transferred from the organization that collected the data to a separate organization that will hold data under the data trust). Finally, there will often be project specific requirements detailed in the documentation used to obtain consent from data subjects, and in the data management plan in submissions to Ethics Boards (REBs or equivalents (e.g., Institutional Review Boards [IRBs]).

Given the focus of this paper is on the sharing of regulated data, such as personal information, that goes beyond uses that data subjects might expect or support, legal authority to collect, hold or share data is critical. Generally, authority will come in the form of at least one, and sometimes more, of the following: (i) authority defined in legislation and/or regulation, (ii) consent on the part of the data subject, (iii) the approval of an REB/IRB [36]. We emphasize legal authority here because we believe that widespread interest in sharing data for public benefit has the potential to result in organizations with good intentions sharing data or providing access to data without having the legal authority to do so.

## Requirement 2: Governance – Four (4) Min Specs

The international and Canadian research literature indicates that members of the mainstream general public are conditionally supportive of data-intensive health research provided that their concerns related to privacy, security and commercial motives are addressed [9,31,37-41]. It is our view that governance is the best way to ensure that data trusts meet all legal requirements AND align with social licence.

Our group identified four min specs related to governance. Foremost, min spec 2a is that a data trust must have a stated purpose. Though the purpose may vary, in our view it is important that the purpose goes beyond the objective of simply sharing data and aims to achieve a specific goal. Further, in the case of data related to publicly funded services, particularly data that are used without express consent [42], we believe that the purpose should include the goal of achieving one or more public benefit(s). For example, a data trust might have the purpose to facilitate the use of person-level data to better understand disease and wellness and evaluate health system interventions.

Data trusts must also have principles regarding how it works towards its purpose. Both the research literature and negative news coverage indicate that transparency with members of the public is particularly important for data sharing [7-12]; therefore, min spec 2b is that the data trust must be transparent in its activities. At a minimum, data trusts should achieve information transparency, e.g., by having plain language information about data holdings and data users available to stakeholders, including members of the public.

To ensure that the purpose and principles are more than words on paper, min spec 2c requires that the data trust establishes a governing body with defined accountabilities. We specify only that there be a body (i.e., not a single person) that is accountable for its decisions and allow that the name and type of the governing body may vary (e.g., Board of Directors, Board of Trustees, Steering Committee), as can the responsibilities and how they are documented (e.g., by-laws, Terms of Reference). As long as the governance and governing body is clear, data trusts could be consolidated under the control of one (lead) organization or a virtual collaborative of partners with mutual interest.

The fourth and final min spec for data trust governance, 2d, is that governance must be adaptive vs. set in stone at the time of establishment. To accomplish this, the responsibilities of the governing body will generally include monitoring for unintended consequences and taking corrective action if activities do not advance the stated purpose or align with principles of the data trust. The governing body will also have to monitor and adapt to changes in legislation and regulation. Without min spec 4d, accountability could decrease over time as the data sharing landscape, risks and opportunities change. The requirement for adaptive governance also increases the likelihood that the data trust will identify and act on "positive risks" such as the emergence of relevant new data sources and new technologies that improve data protection.

## Requirement 3: Management – Three (3) Min Specs

The first data trust management min spec, 3a, "there must be well-defined processes for the collection, storage, use and disclosure of data" encompasses many other more detailed requirements. It is beyond the scope of this paper to describe them all; but fulfilling this min spec will typically involve multiple auditable policies and processes including clear rules around when, how and under what authority data assets are linked or combined. As noted earlier, the Five Safes and other frameworks [1,27-31] can provide guidance on policies and processes for data trust management, and the roles and responsibilities of various actors in the data sharing ecosystem [43,44]. In cases where a data trust involves more than one organization, it is not necessary that all organizations have the same policies and processes. For example, two different data-holding organizations with different de-identification processes might create a joint data trust to link their data, and follow a protocol where the linkage and de-identification is performed using the processes of the organization that contributes the majority of variables to a linked dataset.

While the exact policies and processes can vary, min spec 3b notes that, at a minimum, the policies and processes of 3a must include data protection safeguards which are reviewed and updated regularly. Data protection includes measures to protect against privacy breaches (e.g., unauthorized use of data) and security breaches (e.g., attacks impugning data sovereignty or resulting in loss of control of data). Often data protection safeguards will include validation of user requests including authenticating who the user is, the data required for the scope and reason of use, and the secure environment where use will occur. Compliance monitoring should be in place to identify and respond in cases where there is insufficient protection, unintentional mistakes or deliberate malicious activities. Because threats to privacy and security will change over time, particularly cybersecurity threats, there must be a mechanism to audit privacy and security on a regular basis.

Data trust management's need to be agile and adaptive is not limited to data protection. For example, data trust management bodies need to be aware of, and respond to, new developments in scientific methods/capabilities (e.g., the potential for artificial intelligence and machine learning to provide new insights based on multimodal data), new data sources (e.g., wearables), and changes in public sentiment related to data uses (e.g., in response to news coverage of data breaches). Accordingly, the third management min spec, 3c, is that there is a process in place to identify, assess, track and manage risks. There are many ways to address risk including through policies, administrative processes, data governance, technology, and physical controls. The spirit of min spec 3c is that data trusts need to reassess risks continuously and establish or adapt risk responses as threats and opportunities evolve.

## Requirement 4: Data Users – Two (2) Min Specs

Much of the literature cited above focuses on the responsibilities of data holding organizations; however, it is the data users that are at the frontline of allowable/prohibited activities. To some extent, requirements related to data users are covered by mandatory policies and processes referenced in data trust management min specs. For example, an organization might have a policy that all data users must be vetted bona fide researchers and have the practice of clear provisioning and deprovisioning of data users' rights and access. However, to make it clear that individual data users also have responsibilities, min spec 4a requires individual data users to complete specified training before they access data. The content, length and frequency of the training would be set by the data trust governing body and/or management team and may vary, but the intent of this min spec is to ensure that all data users understand sensitivities associated with the data that they work with, and their obligations related to data use. Therefore, at a minimum, training should educate data users about the limits on how they can use data, e.g., prohibiting attempts to re-identify, barring linkage to other datasets, forbidding the sharing of login credentials, etc.

There are already high quality online materials such as privacy, security and ethics training previously developed by other parties [45] so data trusts would not need to develop all training materials from scratch.

In addition to training, there must be agreements that bind data users, not just the organization(s) that create, manage, and contribute data to the data trust. This is particularly important given growing concern that the processes for de-identification are not foolproof [46,47]. To ensure accountability at the individual data user level, min spec 4b specifies that "all data users must agree to a data user agreement which acknowledges that data use will be monitored and includes consequences for non-compliance." The consequences can vary and may be different depending on the sensitivity of the data. For highly sensitive information, such as health data, consequences such as those that are in included in the UK Biobank material transfer agreement may be appropriate, i.e., in response to non-compliance the data trust 'may prohibit the Applicant Principal Investigator and other researchers from the Applicant's Institution from accessing any further data; and/or, it may inform relevant personnel within the Applicant PI's Institution, funders of the Applicant and/or governing or other relevant regulatory bodies' [48].

## Requirement 5: Public and Stakeholder Engagement – Two (2) Min Specs

Much has been written regarding the importance of public engagement in data-intensive health research, and the importance of doing it well [1,37-41,49-53]. From our perspective there are multiple mechanisms for active and meaningful public and stakeholder engagement, and these may change over the life of the data trust. Given the changing data sharing landscape and heightened public concern about data use, min spec 5a is simply that "there must be early and ongoing engagement with stakeholders including members of the public", i.e., not one-time engagement or engagement that occurs after all decisions have been made.

Further, there is no single "public" [54], and it is not appropriate to rely on mainstream participants to provide the views of groups that are different, for example because they have a disability [55], or a rare disease [56]. For those reasons, min spec 5b notes "Where there is a reasonable expectation that specific subpopulations or groups would have a particular interest in, or would be affected by, an activity of the data trust, there must be direct engagement tailored for that subpopulation/group." In other words, data trusts must supplement the engagement/involvement of the mainstream population with special focused efforts for people with a special stake and concern, in particular, those facing long-standing inequities. Public and stakeholder engagement is not one size fits all.

# Guidance for Implementing the Min Specs

A comparison between this paper and the references we cite would indicate considerable overlap and many consistencies. Our primary contribution is a distillation of ideas and guidance from the literature synthesized with our own experiences to create a relatively short list of min specs for establishing and operating data trusts. In total, we identified twelve practical but essential requirements. We concede that the count is somewhat arbitrary in that we used our judgement to combine min specs that inherently group together and separate those that might be absent from current or planned approaches to data sharing. Notably, though there were technology-related min specs, we did not find that the technological aspects of establishing and operating data trusts present a major challenge. From the perspective of organizations that already are actively working on data sharing, many of the twelve min specs are likely already fulfilled, with some exceptions.

In the case of min spec 2b "The data trust must be transparent in its activities" we are not aware of any organization involved in public sector data sharing that is intentionally opaque. However, with the limited resources available, organizations may not always prioritize work to make information about their activities public and transparent in plain language. We suggest that most organizations could fulfill min spec 2b by adopting an approach similar to the HDR UK Health Data Alliance's requirement to 'publish a register of active projects accessing the data under their custodianship and new data access requests received' since the published information would be also gathered as part of routine data trust operations.

Regarding min spec 4b "All data users must agree to a data user agreement which acknowledges that data use will be monitored and includes consequences for non-compliance" in our experience it is standard practice to have data user agreements signed by researchers and trainees, but there has not been a strong emphasis on individual consequences for non-compliance. If, as the Hall-Presenti report suggests, the goal is widespread sharing and use of data, the future will involve hundreds to thousands of new data users. Among these users, some will make unintentional mistakes and a small subset will be bad actors. In response, we will need consequences for non-compliance that are one step down from the organization level to hold individuals, not just the organizations that they belong to, accountable aligned with the severity and intent of their action.

In the case of min specs 5a and 5b, we find that most organizations with data infrastructure do have some mechanisms for engaging their stakeholders including members of the public; however, it may be treated as a parallel activity vs. one that is integrated into data sharing activities. For example, health data is often collected and shared by hospitals which have patient and family advisory committees. In such cases, it would be a small but necessary step to establish new ongoing mechanisms to inform, consult or involve stakeholders and fulfill min spec 5a. Further, acknowledging that there is no single "public", min spec 5b might require some organizations that are sharing data to go beyond their usual group of advisors, with targeted engagement and involvement for groups and subpopulations with different needs and interests under certain circumstances. We also recommend further engagement to ensure that the labels applied to various forms of data sharing are intuitive and resonate with members of the public.

## Beyond Min Specs

As noted previously, our group's goal was to identify min specs to guide the practical establishment and operations of data trust. During the preparation of this manuscript, several additional elements or characteristics were identified that could strengthen data trusts. These included:
- Dynamic consent for data subjects for data that require consent for collection [57]
- Data traceability so that data trusts can fully execute on patient consent withdrawal, bias monitoring, audits and regulatory agency review [58,59]
- Standard and computable data use conditions [60]
- Secure and auditable computing environments [61]
- Public engagement that goes beyond information transparency and into activities like co-design and deep involvement of data subjects in governance [49-53,62-64]

The fact that these and other potential requirements for data trusts are not included in our list of min specs (Box 1) does not mean that they are unimportant add-ons. It is possible that these and other requirements become the norm as threats and opportunities related to data sharing increase, and as technological approaches to data protection mature and become more widespread. Our stringent criteria regarding what constitutes a min spec stems from our first-hand experience with data infrastructure. In practice, it is necessary to find a balance between totally locked-down data and/or extensive technological control of data with ease of use and the cost to establish and maintain data infrastructure. Even for light-touch governance and management for non-sensitive data, there still needs to be funded staff to ensure the provisioning of users, security protocols, public engagement etc. Data infrastructure, especially distributed data infrastructure, may not have the look of traditional large-scale research infrastructure like wet labs, large microscopes and other scientific equipment, but it still needs to be funded. Accordingly, we have identified the min specs that we believe are essential requirements with the hope that focusing the available funding on them will enable the most, and the most responsible, data sharing possible with resources available.

## Limitations

Foremost, this paper presents the views and recommendations of representatives from Canadian publicly funded data-related initiatives, with a focus on health data. It may not reflect the views and data sharing activities of other organizations/initiatives in Canada and in other jurisdictions. Secondly, the list of min specs has not yet been discussed and refined with members of the public. Additional work is planned in that regard.

Finally, the paper includes only minimal input from commercial organizations which are active in the field of data sharing and data security.

# CONCLUSIONS

We identified a relatively small number (12) of min specs for establishing and operating data trusts which should be practical to implement. The mechanism of a capability exchange combined with min specs facilitation was effective for identifying essential requirements for data trusts. This feature paper is just a start; continued joint work with members of the public, representatives from commercial organizations and other Canadian and international organizations involved in data infrastructure is recommended on this evolving topic.

# ACKNOWLEDGEMENTS

The authors thank Alexandre Le Bouthilier, co-lead of the Terry Fox Research Institute/Imagia Digital Health and Discovery Platform funded by the Canadian government, who could not attend the December meeting but contributed to this manuscript.

# STATEMENT OF CONFLICTS OF INTEREST

No conflicts of interest were identified. Two individuals have significant involvement in commercial organizations. Brian Courtney (co-author) has employment, significant ownership, royalties, and a director position related to Conavi Medical Inc. Alexandre Le Bouthilier (acknowledged contributor) is cofounder and shareholder of Imagia.

# ETHICS STATEMENT

This manuscript did not involve primary research on human subjects and was not submitted for Research Ethics Board approval.

# ABBREVIATIONS

AI: artificial intelligence

EHDEN: European Health Data & Evidence Network

FAIR: findable, accessible, interoperable, reusable

FIPPA: *Freedom of Information and Protection of Privacy Act*

HDR UK: Health Data Research UK

ICES: Institute for Clinical Evaluative Sciences

IJPDS: International Journal of Population Data Science

IRB: Institutional Review Boards

ML: machine learning

NHS: National Health Services

OECD: Organization for Economic Co-Operation and Development

PI: Principal Investigator

PIPEDA: *Personal Information Protection and Electronic Documents Act*

PHIPA: *Personal Health Information Protection Act*

REB: Research Ethics Boards

UK: United Kingdom